\documentclass[conference]{IEEEtran}
\IEEEoverridecommandlockouts
\usepackage{cite}
\usepackage{amsmath,amssymb,amsfonts}
\usepackage{algorithmic}
\usepackage{graphicx}
\usepackage{hyperref}
\usepackage{textcomp}
\usepackage{url}
\usepackage{xcolor}
\def\BibTeX{{\rm B\kern-.05em{\sc i\kern-.025em b}\kern-.08em
    T\kern-.1667em\lower.7ex\hbox{E}\kern-.125emX}}
\begin{document}


\makeatletter
    \newcommand{\linebreakand}{%
      \end{@IEEEauthorhalign}
      \hfill\mbox{}\par
      \mbox{}\hfill\begin{@IEEEauthorhalign}
    }
    \makeatother

\title{Beyond Certificates: 6G-ready Access Control for the Service-Based Architecture with Decentralized Identifiers and Verifiable Credentials\\
\thanks{This work has been submitted to the IEEE for possible publication. Copyright may be transferred without notice, after which this version may no longer be accessible.}
}

\author{\IEEEauthorblockN{Sandro Rodriguez Garzon}
\IEEEauthorblockA{\textit{Service-centric Networking} \\
\textit{Technische Universität Berlin/T-Labs}\\
Berlin, Germany \\
sandro.rodriguezgarzon@tu-berlin.de}
\and 
\IEEEauthorblockN{Hai Dinh-Tuan}
\IEEEauthorblockA{\textit{Service-centric Networking} \\
\textit{Technische Universität Berlin/T-Labs}\\
Berlin, Germany \\
hai.dinhtuan@tu-berlin.de}
\and
\IEEEauthorblockN{Maria Mora Martinez}
\IEEEauthorblockA{\textit{Service-centric Networking} \\
\textit{Technische Universität Berlin/T-Labs}\\
Berlin, Germany \\
m.moramartinez@tu-berlin.de}
\and
\linebreakand 
\IEEEauthorblockN{\hspace{-0.8cm}Axel Küpper}
\IEEEauthorblockA{\hspace{-0.8cm}\textit{Service-centric Networking} \\
\hspace{-0.8cm}\textit{Technische Universität Berlin/T-Labs}\\
\hspace{-0.8cm}Berlin, Germany \\
\hspace{-0.8cm}axel.kuepper@tu-berlin.de}
\and
\IEEEauthorblockN{Hans Joachim Einsiedler}
\IEEEauthorblockA{\textit{Group Technology} \\
\textit{Deutsche Telekom AG}\\
Berlin, Germany \\
hans.einsiedler@telekom.de}
\and 
\IEEEauthorblockN{Daniela Schneider}
\IEEEauthorblockA{\textit{Core \& Network Services} \\
\textit{Deutsche Telekom AG}\\
Vienna, Austria \\
daniela.schneider@magenta.at}}


\newcommand{\ts}{\textsuperscript}

\maketitle

\begin{abstract}

Next generation mobile networks are poised to transition from monolithic structures owned and operated by single mobile network operators into multi-stakeholder networks where various parties contribute with infrastructure, resources, and services. However, a federation of networks and services brings along a crucial challenge: Guaranteeing secure and trustworthy access control among network entities of different administrative domains. This paper introduces a novel technical concept and a prototype, outlining and implementing a 5G Service-Based Architecture that utilizes Decentralized Identifiers and Verifiable Credentials instead of  traditional X.509 certificates and OAuth2.0 access tokens to authenticate and authorize network functions among each other across administrative domains. This decentralized approach to identity and permission management for network functions reduces the risk of single points of failure associated with centralized public key infrastructures. It unifies access control mechanisms and lays the groundwork for lesser complex and more trustful cross-domain key management for highly collaborative network functions in a multi-party Service-Based Architecture of 6G.


\end{abstract}


\section{Introduction}

In recent years, public landline mobile networks (PLMNs) have transformed from monolithic structures to multi-stakeholder networks, ushering in an era where concepts such as OpenRAN, network slicing, network and service federation, mobile edge computing, and infrastructure sharing will become reality and dominate the architecture discussions about the next mobile network generation 6G \cite{An.2021}. Network entities engaging in cross-domain interactions even within PLMNs will increasingly become the standard rather than the exception \cite{Bertin.2021}. As the evolution unfold, the role of access control within the access and core networks will shift from being optional to being indispensable for secure and trustful networking. 

Within the Service-Based Architecture (SBA) of the 5G core network's control plane, access control utilizes Transport Layer Security (TLS) with X.509 certificates according to Annex E of TS 33.310 for the authentication of network functions (NF). In addition, the authorization can be conducted with access tokens according to the OAuth2.0 industry standard. An X.509 certificate is used primarily to verify the identity of an NF that holds the private key associated with the certificate's public key. It is issued by a trusted certificate authority (CA) as part of an X.509 public key infrastructure (PKI). However, due to a lack of a single globally trustworthy CA for all PLMNs, each mobile network operator (MNO) is assumed to operate its own PKI with at least one centralized CA and to apply cross-certification among CAs of different PLMNs to enable cross-domain authentication of NFs in a roaming scenario \cite{GSMAssociation.2021}. This approach, while effective to a certain degree, brings about significant vulnerabilities. The primary concern is that each CA becomes a single point of failure, jeopardizing potentially the security and reliability of the entire ecosystem \cite{Won:2018}. Furthermore, as the number of stakeholders grows, the intricacy of managing interconnected PKIs surges, posing scalability, complexity, and as a result efficiency challenges.

Decentralized Identifiers (DIDs) \cite{WorldWideWebConsortium.822021}, in contrast, are an attempt to standardize the way identifiers and verification material associated with a subject that is being referred to by the identifier are managed in a decentralized fashion. An entity equipped with a DID becomes its own identity provider. This eliminates the need for a CA-issued X.509 certificate, which typically encompasses an identifier that is determined and controlled by the CA. Verifiable Credentials (VC) \cite{WorldWideWebConsortium.05.11.2021} are then tamper-resistant claims that are cryptographically bound to a DID, issued by a trustful 3\ts{rd}-party, and hold by the identity subject itself. A VC enables an entity to prove identity and arbitrary sorts of claims towards others in a cryptographically verifiable way without actively involving the VC's issuer. Since DIDs and VCs are self-managed and the former, in addition, being self-issued there is no need to operate full-fledged PKIs with potentially vulnerable CAs, nor to interconnect them for cross-certification purposes.


This paper is an initial attempt to conceptualize the use of DIDs and VCs for access control purposes within the 5G's SBA. As a first proof of concept, it introduces an SBA which assigns DIDs to NFs and encodes identity claims and permissions grants of NFs as DID-anchored VCs. The objective is to transition away from complex X.509 certificate management for access control purposes, which currently involves at least a single CA per stakeholder of a 5G ecosystem. Instead, this work aims to unify and make access control in the SBA multi-stakeholder-ready by the introduction of decentralized identity and permission management for NFs, charting a path towards more secure, trustful, efficient, and highly-collaborative PLMNs in 6G.

\section{Preliminaries}

\subsection{Service-Based Architecture}

In contrast to former mobile network generations, the control plane of the 5G core network has been designed as a Service-based Architecture (SBA), to incorporate state-of-the-art Internet technologies and service design patterns. The 5G SBA consists of different independent modules or services that are denoted as Network Functions (NF) as described in TS 23.501. Each NF must expose a Service-based Interface (SBI) which offers a RESTful API so that other NFs (in the role of consuming NFs) can interact with and thus consume services offered by the NF (in the role of a producing NF). Access to the SBI is conducted with HTTP/2 and JSON on the application layer, as defined in TS 29.500. According to TS 33.501, TLS on the transport layer can optionally be supported by NFs for security protection if network protection is not provided by other means such as with IPSec on the network layer. NFs are equipped with X.509 certificates so that they are able mutually authenticate themselves during a TLS handshake. Neither the management of X.509 certificates for NFs with PKIs nor the management of trust relationships among issuers and verifiers of X.509 certificates is in the scope of the 5G specification. In addition, TS 33.501 specifies authorizations to be conducted optionally with OAuth2.0 on the application layer. The network repository function (NRF) takes over the role of the authorization server besides being a service discovery server, provisioning consuming NFs with OAuth2.0 access tokens before they access a service of a producing NF. The Client Credentials Flow grant type, specified in clause 4.4 of IETF RFC 6749, is applied here whereby a client ID and a client secret is required by a consuming NF for an access token to be granted by the NRF.

\subsection{Decentralized Identity Management}
A digital identity is a set of electronically stored attributes, such as a name or type, that uniquely represents an identity subject, such as a human being or a thing. It serves as a digital counterpart to a subject's real-world identity and provides the basis to establish trust among different actors in the digital domain. In centralized and federated Identity Management (IDM) systems, identity data is owned and controlled by identity providers, not the identity subjects. With decentralized IDM, the latter have ownership and full control over their identity data. However, this demands a standardized and tamper-resistant method to encode identity claims.

\begin{figure}[!t]
\centerline{\includegraphics{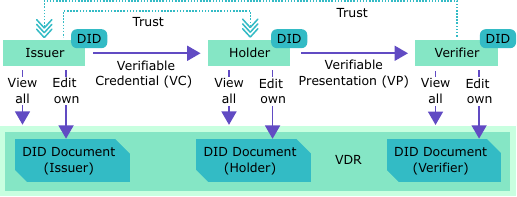}}
\caption{Issuance and presentation of DID-bound verifiable credentials.}
\label{fig:ssi}
\end{figure}


The Verifiable Credentials (VC) model is a first attempt to conceptualize and standardize a tamper-proof container that stores the identity attributes in the form of claims. A VC is not issued by the identity subject itself, but by an issuer who verifies the claims about the identity subject and confirms their integrity by generating a proof, such as a digital signature, which becomes part of the VC. Once the VC is created, the issuer hands it over to the identity subject, who is now in control of the VC and can decide where and how it is stored and with whom to share it. A verifier, with whom a VC is shared by an identity subject within a Verifiable Presentation (VP), e.g., for authentication purposes, can verify its authenticity and integrity. But VPs and contained VCs require unique references to the parties involved in its creation, including the issuer and the identity subject the claims are about. This is accomplished with DIDs. A DID is a unique identifier that refers to an entity and resolves to a DID document. The latter contains public key material so that the entity that is referred to by the DID can proof ownership to others by means of the underlying private key material. DID documents are accessed through a Verifiable Data Registry (VDR), which is a tamper-proof, preferably decentralized infrastructure, e.g., a distributed ledger (DL). Only the DID owner can edit its DID document in the VDR but view all others. DIDs are intended to be created, owned, and controlled by the entity they refer to. An issuer, an identity subject holding a VC (denoted as holder), and a verifier each owns a unique DID. During creation, VCs are cryptographically bound by the issuer to the DID of the identity subject the claims are about. A VP then contains the VC and is, in addition, digitally signed by the identity subject. As a result, a VP empowers the identity subject to proof ownership over its DID and to present tamper-resistant and 3\ts{rd}-party issued verifiable claims about the entity the DID refers to, the identity subject itself. Fig. \ref{fig:ssi} illustrates how the issuer, holder, and verifier are related to each other.

Traditional X.509 certificates mix the concepts of identity and key management; they are containers that bind an identity to its key material, secured by the CA's signature. DIDs and VCs, in contrast, clearly separate identity attributes in VCs from key material in the DID documents and link them by the DID. This facilitates not only the realization of fundamental security features such as authentication, integrity, and confidentiality, but also advanced functions such as authorization and assertion, which X.509 is originally not designed for. VC formats even provide advanced mechanisms to protect the VC's attributes that are not supported in X.509, including selective disclosure and zero-knowledge proofs.

\section{Related Work}

The inherent characteristic of a DID is that once an entity owns the underlying key pair it can prove ownership of it to others. With DID-bound VCs, an entity can even clearly identify itself towards others through tamper-proof and verifiable claims. Since DIDs and VCs were originally made to represent human beings and their claims, there exist numerous approaches about how the authentication procedure can be conducted with a human in the loop \cite{Sabadello:2018}, e.g., by typically involving a human to scan QR codes \cite{Sabadello:2018, Lux:2020, Yildiz:2021}. In the context of the SBA, however, entities in form of software or hardware service instances need to mutually authenticate themselves without the intervention of a human. It reassembles an Internet of Things (IoT) environment, where things do not only mutually authenticate themselves autonomously but also need to handle permissions for authorization purposes. 

An access control model based on DIDs and VCs that gets along without human intervention was developed within the QualiChain project \cite{Belchior:2020}. In the proposed model, which is capable to implement role- as well as attribute-based access control, agents act autonomously on behalf of humans and present VPs in order to access resources. For constraint IoT devices, Lagutin et al. present a method to use DIDs for authentication purposes and VCs for authorization purposes within the ACE-OAuth flow, an OAuth2.0 framework for constraint environments \cite{Lagutin:2019}. In \cite{Jung:2021}, DIDs are used to authenticate an entity towards another entity while access policies are encoded within the DID document of the resource provider. However, the latter extensions violate the DID specification and contradicts the primary intention of a DID document to contain only pseudo-anonymous verification material of an entity. Kim et al. proposes DIDs and VCs for identification while the authorizations are given through local policies by a resource provider. In \cite{Saidi:2022}, DIDs and VCs are used for authentication and the role-based permission policies are stored in a smart contract, which decentralizes the policy decision point.        

In the realm of mobile networks, the concepts and ideas that DIDs and VCs are based upon are primarily applied for the subscriber authentication. You et al. discusses the potential and benefits of sharing verification material for subscribers in 5G via a DL in order to efficiently authenticate visiting subscribers \cite{Yue:2021}. Although not using DIDs and VCs, Xu et al. describe an approach to enable subscriber authentication with the help of locally stored verifiable claims (similar to VCs) and verification material shared via a DL (similar to DL-anchored DID documents)\cite{Xu:2020}. Haddad et al. follows a similar path and introduces a new authentication and key agreement protocol for 5G that authenticates visitors in 5G networks without the need to consult the home PLMN authentication center \cite{Haddad:2020}. From a standardization point of view, the 5G Public-Private-Partnership (5GPPP) started to emphasize the general challenge of secure and trustful cross-domain access control and describes DIDs and VCs as potential key concepts to conduct trustful cross-domain identity and permission management in 5G and beyond \cite{5GPPP:2021}. Rodriguez Garzon et al. then adds access control of network entities within the core networks to the range of application areas in 6G that can benefit from the usage of DIDs and VCs\cite{Garzon:2022}. Within the 5GZORRO project, Valero et al. introduce a security and trust framework for 5G networks to let entities of different operators create secure VPN tunnels among them by means of the verification material found in each other's DID documents within a DL \cite{Valero:2021}. Despite mentioning the use of VCs for the identification and authorization of stakeholders across administrative domains, it remains unclear what role VCs play in the setup of a VPN channel. But besides sketching the idea of a DID-based decentralized PKI for the SBA\cite{Garzon:2022} and using VCs for the authorization of NFs within the SBA\cite{Garzon:2022b}, so far, there exists no technical concept of DID- and VC-enabled access control for the SBA nor a proof of concept. 



\section{Concept}

In an SBA where access management utilizes DIDs instead of X.509 certificates and VCs instead of OAuth2.0 access tokens, each NF must possess at least a DID. An NF owns multiple DIDs if it intends to use different digital identities to interact with different NFs. It can either create the DID itself by generating a key pair or letting an entity of the management plane provision it with a DID including the underlying key pair. An NF's DID can be of public or private nature. With a public DID, the DID document is anchored respectively stored in a VDR. With a private DID, also referred to as a peer DID, the DID document can be extracted from the DID itself. The type of DID to use for NFs depends primarily upon the need to subsequently change the DID document. The latter of a private DID can't be changed while its public counterpart can subsequently be adapted  by the DID owner in a cryptographically verifiable way within the VDR. Private DIDs are therefore better suited for short-lived NFs while their counterparts are a better fit for long-living NFs, e.g., that rotate keys on a regular basis to enhance security. 

A DID-equipped NF, regardless of the DID being public or private, can authenticate itself only pseudo-anonymously towards other NFs because the DID document does not reveal the identity of the referred entity. To clearly identify itself towards others, the NF needs, in addition, verifiable identity claims in the form of VCs that are bound to the NF's DID. For example, a VC might state that an NF is of a specific type, belongs to a network slice, or is owned by a specific MNO. These type of VCs form the basis of an NF's digital identity and are in the following denoted as \textit{AuthN VCs}. They are used by NFs during access control to identify themselves towards others in addition to the pseudo-anonymous authentication with a DID. For the authorization, the authorizer can either give permission to access its resources based on the verifiable identity claims in the AuthN VCs of the requester (attributed-based access control), the role of the requester as part of the same AuthN VCs (role-based access control), or in the spirit of an access token through a VC encoding the permission grant. The former two options can be realized with local policies at the requested NF or by a dedicated and trustful policy decision function as proposed in \cite{Belchior:2020} or \cite{Saidi:2022}. With the latter option, the requester is granted permission solely based on a presented VC, here denoted as an \textit{AuthZ VC}, which contains the access permission as verifiable key/value pairs. 

While DIDs are self-issued, regardless of being self-created or handed over by the management plane during NF instantiation, verifiable identity attributes for identification (AuthN VCs) and verifiable permissions grants (AuthZ VCs) for authorization must be issued by a trusted 3\ts{rd}-party. Consequently, there is a need for trustful DID-enabled entities in the SBA that act in the role of identity providers, issuing AuthN VCs, and authorization servers, issuing AuthZ VCs. Although being logically different type of entities, they are subsumed here under the term \textit{identity and permission management function (IPMF)}. Besides the issuance of VCs, IPMFs are also responsible to revoke VCs, e.g., once an NF is compromised or the access permission has been withdrawn. The revocation mechanism can thereby be implemented via a DL, so that a verifying entity is able to check the revocation status of a VC by inquiring the DL. Alternatively, an IPMF can provide a VC with a validity period and thus, if the security policies permit, eliminate the need for a revocation mechanism.

Once NFs are equipped with DIDs and DID-bound AuthN and AuthZ VCs, they can (mutually) identify and authorize themselves. Identification subsumes a) the pseudo-anonymous authentication with DIDs and b) the identification with AuthN VCs. While a) can alternatively be conducted on the transport layer as part of a modified TLS handshake \cite{Perugini:2023} or with a VPN tunnel \cite{Valero:2021}, or on the application layer as part of an application layer transport protocol such as DIDComm\cite{DIDComm:2020}, b) happens only on the application layer with a VP exchange protocol such as the DIF Presentation Exchange \cite{DIF.2022}. The latter is also applicable to the authorization with AuthZ VCs since they are technically alike and only differ in the intended purpose. AuthN and AuthZ VCs can even be combined into a single VP, unifying and simplifying access control procedures.  

DID and VC-based access control happens in a uni- or bidirectional manner, depending on the scenario-specific requirements. During access control, an NF A (verifier) considers a VP (comprising VCs) presented by a NF B (holder) as technically valid as long as 1) the VCs can be verified with the verification material of the IPMFs (issuer), 2) the VP as a whole can be verified with NF B's verification material, and, optionally, 3) the VCs are not revoked by the IPMFs. The required verification material is provided by the VDR, if NF B and the IPMFs own public DIDs, or extracted from the DIDs itself, if private DIDs are used. The extent to which the identity or permissions claims of NF B are trustworthy, however, depends upon NF A's trust in the involved IPMFs. 


\begin{figure}[!t]
\centerline{\includegraphics{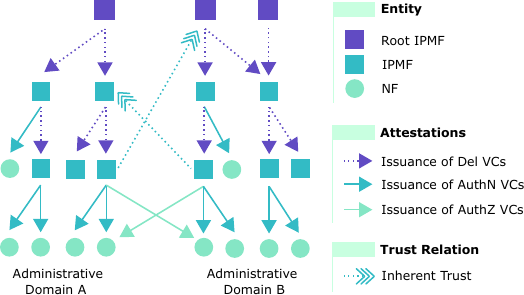}}
\caption{Exemplary domain-specific issuer hierarchies assembled through right delegations and exemplary identity and permission attestations for NFs.}
\label{fig:ipmf-nf}
\end{figure}

Multiple IPMFs within the same administrative domain can be structured hierarchically by enabling IPMFs (denoted as parent IPMFs) to delegate subsets of their rights to other IPMFs (denoted as child IPMFs). These delegable set of rights must always include the right to issue AuthN and/or AuthZ VCs to NFs and might also include the right to subdelegate the former rights to another IPMF. Delegations are encoded as VCs, denoted as Del VCs, and are issued by each parent IPMF to its child IPMFs. Del VCs received from a parent IPMF also contain the Del VCs the parent IPMF received from its parent as so forth. This empowers an IPMF to cryptographically prove to others that the delegations it received are valid all the way up to the first Del VC issued by the root IPMF. Since an NF receives the chain of Del VCs of its IPMF as part of the AuthN/AuthZ VCs, a verifying NF can securely trace back a VP to an IPMF which is well-known and trusted by it. Del VCs enable organizations to align a hierarchy of IPMFs to their organizational layout, to allocate the operational issuance tasks among different entities for optimal load sharing, and to distinctly protect the critical upper IPMFs, e.g., the root IPMF, by letting them only delegate and not participate in the operational issuance process for potentially harmful external NFs. Fig. \ref{fig:ipmf-nf} illustrates how a exemplary trust hierarchy of IMPFs is established by the issuance of Del VCs and where NFs can potentially receive their identity claims (AuthN VCs) from. It also shows that when an NF receives a cross-domain permission grant, the issuing IPMF needs a fundamental level of trust in at least one IPMF within the trust chain of the NF's administrative domain. Otherwise, an AuthN VC provided by the NF during the identification phase of the issuance process can't be reliably traced back to a trustworthy IPMF.


\begin{figure*}[ht!]
\centerline{\includegraphics[scale=1.0]{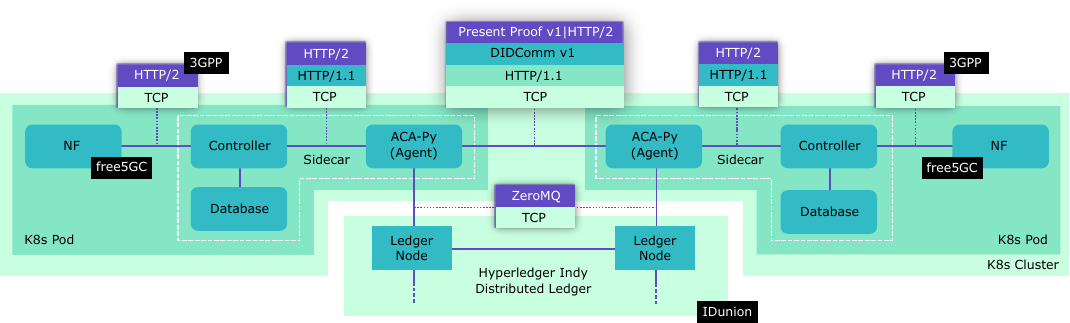}}
\caption{NF-sidecar architecture with DIDComm-tunneled HTTP/2 communication link between NFs.}
\label{fig:sidecar}
\end{figure*}

In highly dynamic cloud environments, virtualized NFs are numerous and might come and go while IPMFs are fewer in number and are supposed to exist for a longer period, in particular the root IPMF as the major trust anchor. This makes an NF a good candidate to own private DIDs especially if edit operations of a DID document in a VDR are costly, e.g., in a DL, and as long as a key rotation can be omitted. IPMFs, on the other hand, are critical long living entities where at least a periodical change of verification material in the DID document is mandatory from a security point of view. IPMFs are therefore predestined to own public DIDs, especially a root IPMF, because their DID documents can not only be adapted subsequently but be anchored in a DL that is commonly operated and equally governed by multiple stakeholders and is therefore well-suited to become a global common source of truth for verification material of major trust anchors.



\section{Implementation}

For the prototypical DID- and VC-enabled SBA, several cloud-native solutions were integrated to closely follow a state-of-the-art core network deployment model. As a foundation, the open-source framework free5GC\footnote{https://free5gc.org/} was used; a fully operational 5G core network in alignment with 3GPP Release 15. The User Equipment (UE) and the Radio Access Network are simulated with UERANSIM\footnote{\url{https://github.com/aligungr/UERANSIM}}. All NFs, the IPMF, the gNodeB, and the UE are containerized and orchestrated in a Kubernetes (K8s) cluster. An Aries Cloud Agent\footnote{\url{https://github.com/hyperledger/aries-cloudagent-python}} (ACA-Py) at each NF manages DIDs and VCs, integrates a digital wallet, interacts with the VDR, and enables message-based and transport-agnostic communication among the agents of different NFs with DIDComm\cite{DIDComm:2020} (v1). DIDcomm is an application layer transport protocol for the realization of secure end-to-end communication channels with DIDs, ensuring confidentiality, integrity, and authenticity. Each NF and the IPMF runs in a separate K8s pod and owns a unique public DID that is anchored in the VDR. The entire NF communication logic, incl. access control, is encapsulated in a sidecar, encompassing three entities. The controller, an interface tailored for the ACA-Py agent's API, is the sidecar's central hub. The agent is the interface towards other NF sidecars and the VDR while the database keeps state of the associations among NF sidecars. The architecture and protocol stacks are illustrated in Fig. \ref{fig:sidecar}. The DL of the IDunion project\footnote{\url{https://idunion.org}} serves as the VDR and is powered by Hyperledger Indy\footnote{\url{https://www.hyperledger.org/projects/hyperledger-indy}}. It is commonly operated by the IDunion project members, coming from various economic sectors such finance, transportation, telecommunication, and manufacturing. A legal entity of a European cooperative, a Sociedad Cooperativa Europea (S.C.E), was founded in IDunion as a legal framework for the governance of the DL.     




In the prototype, each NF is initially shipped with a DID and a basic set of rudimentary AuthN VCs. These pre-configured VCs are utilized by the NFs during the issuance process to initially identify themselves towards the primary issuing party, the IPMF. After they successfully received AuthN and AuthZ VCs from the IPMF for operational purposes, they are ready to identify and authorize themselves towards other NFs and IPMFs during operation, even across administrative domains. The issuance of VCs by the IPMF and the mutual presentation of VCs by NFs for access control purposes are conduced on top of the simplex DIDComm protocol. Each DIDComm message's payload is hereby encrypted with a symmetric key which is in turn encrypted by the sender with the sender's private key and the receiver's public key (as known from the receiver's DID document) and attached to the message. Only the intended receiver is able to decrypt the symmetric key and thus the message payload, and to verify the message's integrity and authenticity. A mutual identification is conducted between all NFs, incl. the NRF. For authorization, a consuming NF needs to present AuthZ VPs to the producing NF but not vice versa. Access control through the exchange of AuthN and AuthZ VPs happens only once between a pair of NFs. VCs are encoded as AnonCreds\footnote{\url{https://hyperledger.github.io/anoncreds-spec/}}, issued by an IPMF using the Aries Issue Credential Protocol\footnote{\url{https://github.com/hyperledger/aries-rfcs/blob/main/features/0036-issue-credential/README.md}}, and are presented in VPs by an NF for access control purposes with the Aries Present Proof Protocol\footnote{\url{https://github.com/hyperledger/aries-rfcs/blob/main/features/0037-present-proof/README.md}}. Fig. \ref{fig:sequence-auth} illustrates the combined mutual identification and one-way authorization procedure among NFs of the prototype. It demonstrates how access control procedures can be unified due to the flexibility of VCs and the ability of a VP to carry multiple VCs that may serve different purposes and originate from different issuers.




\begin{figure}[t!]
\centerline{\includegraphics[scale=0.95]{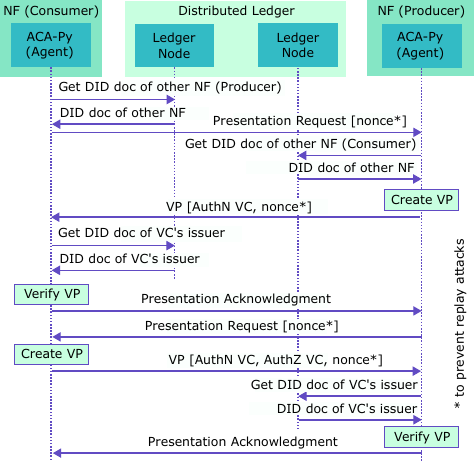}}
\caption{Combined mutual identification and one-way authorization among NFs through the presentation of AuthN and AuthZ VCs.}
\label{fig:sequence-auth}
\end{figure}

\begin{figure*}[ht!]
\centerline{\includegraphics[scale=0.94]{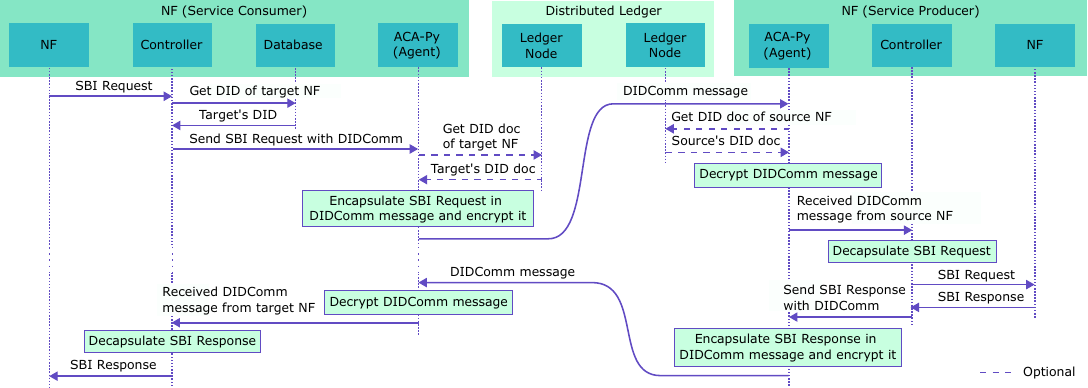}}
\caption{Exemplary call of an NF's SBI (service producer) by another NF (service consumer) through their sidecars via the secured DIDComm channel.}
\label{fig:sequence}
\end{figure*}


The 3GPP-compliant free5GC implementations of the NFs are adopted in the prototype without change. However, the NF instances are configured in such a way so that all in- and outgoing communication comes and goes through their sidecars. So all SBI-specific HTTP/2 traffic is tunneled through the pairwise DIDComm channels that are managed by the sidecar. If an NF initiates a request to another NF, it will be intercepted by the sidecar's controller. The latter determines the targets NF's DID from the information gained from the request. In the prototype, the derivation of the target's DID from the request's meta data is pre-configured because extending the NF Profile to contain a DID in addition to the URL would require an adaptation of the NRF and service discovery response, resulting in an incompatibility with Release 15. In a future SBA in 6G, the NRF is assumed to provide this missing piece of information as part of the NF Profile, rendering the derivation process obsolete. The controller will forward the request encapsulated in an HTTP/1.1 message and enriched with the target DID to the sidecar's agent. The agent encapsulates the original HTTP/2 request into a DIDComm message, encrypts it, and forwards it to the target's agent. Whether or not a latest version of the receiver's DID document need to be retrieved from the DL prior to the encryption processing step depends upon the security policies. If the cryptograhic material of a receiver's DID document is not assumed to change during the existence of a DIDComm channel then no update of a DID document is needed by the sender. Otherwise, it needs to query the DL on a regular basis or gets proactively notified in case the receiver's DID document changes. At the target NF's sidecar, the request goes through a reverse process and is then forwarded in a decapsulated form to the target NF. In the reverse direction, the response goes through similar processing and forwarding stages. However, the DID of the requesting NF does not need to be resolved because it becomes known by the target's sidecar through the request. Fig. \ref{fig:sequence} illustrates an exemplary DIDComm-tunneled SBI call. The decision to go with a slightly lesser performant tunnel solution was driven by the requirement to be fully compatible with the current 5G SBA implementations of the SBI (legacy systems compliance). However, this makes the sidecar solution future-proof for upcoming iterations of the free5GC or similar 5G core implementations.

For an initial assessment of the performance, a UE registration as defined in TS 23.502 and encompassing 58 REST calls incl. NRF service discovery requests among six NFs, was conducted in an experimental setup on a virtual machine hosted on a server powered by an AMD EPYC 7262 CPU, with 4 vCPUs, 16GB of memory, and Ubuntu 22.04.2 LTS with kernel version 5.15.0-73-generic. A complete UE registration in the original and unaltered SBA takes on average 5.129 sec. without TLS and 5.130 sec. with TLS; after 30 iterations. With a DID- and VC-enabled SBA that relies on DIDComm for the transport, the duration extends to 5.838 sec on average; about a 13.8\% increase. Since the DID documents were locally cached during the evaluation it does not include the time to retrieve each others DID documents from the DL.

\section{Conclusion}
This work presented a first concept and a prototype of an SBA where NFs utilize DIDs and VCs as an alternative to X.509 certificates and OAuth2.0 access tokens to authenticate and authorize each other. As a proof of concept, it demonstrates the technical feasibility to eliminate the need for centralized and vulnerable CAs within full-fledged PKIs for the SBA without having to sacrifice core security features related to them such as verifiable digital identities and permissions grants or secure communication channels. In fact, the proposed approach to decentralize identity and permission management for NFs in the SBA comes along with benefits, such as simplified and trustful cross-domain key management, unified access control as well as improved security against centralized breaches. The introduction of IPMFs for the issuance and, optionally, the revocation of verifiable identity and permission claims empowers NFs during operation to adapt on-demand to evolving access control requirements in highly dynamic multi-stakeholder environments without requiring full CI/CD pipeline runs. Furthermore, it relieves the NRF from acting as an (cross-domain) authorization server for NFs by logically outsourcing permission management to a dedicated network entity. Currently, however, it lacks standardized technical means to setup highly efficient and secure connection-oriented channels with the help of DIDs on the transport layer, which is assumed to have a significant impact on the performance of transported HTTP/2 REST calls. With respect to a commonly operated and most probably permissioned DL to share DID documents of major trust anchors for access control purposes in the SBA, it remains still unclear how it will be governed by stakeholders of a global 6G ecosystem in which geopolitical, market, and regulatory interests collide.

\section*{Acknowledgment}
We would like to thank Artur Philipp from the Technische Universität Berlin for his valuable advice and support. This work was funded by the Deutsche Telekom AG. 

\bibliographystyle{IEEEtran}  
\bibliography{references}  

\end{document}